\documentclass[11pt,twoside]{article}

\usepackage{asp2006}
\usepackage{epsf}
\usepackage{psfig}
\usepackage{lscape}

\begin{document}

\title{Shell Galaxies, Dynamical Friction, and Dwarf Disruption}   

\author{
I. Ebrov\'a,$^{1,2}$
B. Jungwiert,$^2$
G. Canalizo,$^3$
N. Bennert,$^4$
and L. J\'ilkov\'a$^5$
}

\affil{
$^1$Faculty of Mathematics and Physics, Charles University in Prague, Ke~Karlovu~3, CZ-121~16 Prague, Czech Republic\\
$^2$Astronomical Institute, Academy of Sciences of the Czech Republic, Bo\v{c}n\'{i} II 1401/1a, CZ-141 31 Prague, Czech Republic\\
$^3$IGPP \& Dept. of Phys.,\,Univ.\,of California,\,Riverside,\,CA 92521,\,USA\\
$^4$Dept. of Physics, Univ. of California, Santa Barbara, CA 93106, USA\\
$^5$Dept. of Theoretical Physic and Astronomy, Faculty of Science, Masaryk University, Kotl\'a\v rsk\'a 2, CZ-611 37 Brno, Czech Republic
}

\begin{abstract}
Using N-body simulations of shell galaxies created in nearly radial 
minor mergers, we investigate the error of collision dating, 
resulting from the neglect of dynamical friction and of
gradual disruption of the cannibalized dwarf.   
\end{abstract}

We compared a simulation without dynamical friction and with instant disruption of the elliptical dwarf during the first passage through the center of a giant elliptical (run 1) to a model with the same initial conditions but dynamical friction and gradual decay of the dwarf involved (run 2). Only position of the outermost shell remains almost unaffected but its brightness is drastically lowered. If we observationally identified the second outermost shell to be the outermost one, we would underestimate the merger age by several Gyr. For details and references, see \citet{usa}.

\begin{figure}[!ht]
\plotone{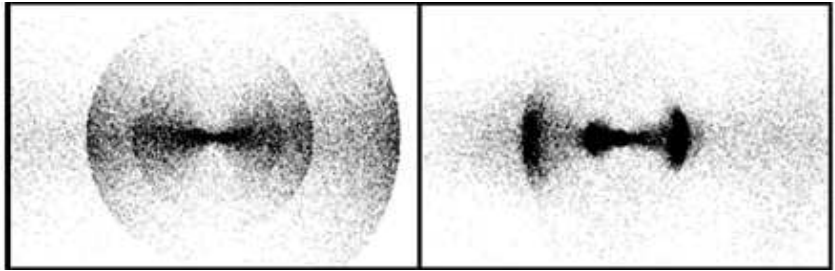}
\caption{
Snapshots of simulations -- run 1 (left) and run 2 (right) -- at 4.5 Gyr after beginning of the merger. Only stars of the dwarf are shown. Each box, centered on the host galaxy, shows 300$\times$200 kpc.
}\label{fig}
\end{figure}

\acknowledgements This project is supported by grants AV0Z10030501 (Acad. of 
Sci. of the Czech Rep.), 205/08/H005 (Czech Science Foundation) and LC06014 
(Czech Ministry of Education).


\begin{thebibliography}{}
\bibitem[Ebrov\'a et al.(2009)]{usa}
Ebrov\'a, I., Jungwiert, B., Canalizo, G., Bennert N., J\'\i lkov\'a, L., 2009,
arXiv0908.3742
\end{thebibliography}
\end{document}